%Paper: hep-th/9303160
%From: "Victor Gurarie" <gurarie@puhep1.Princeton.EDU>
%Date: Mon, 29 Mar 93 18:02:49 EST
%Date (revised): Thu, 8 Apr 93 23:40:53 EDT

\font\title = cmr10 scaled \magstep 2
\font\subtitle = cmr10 scaled \magstep 1

\magnification=\magstep 1

{ \centerline{}
{ \bf March 1993 \hfill PUPT-1391}
\vskip 50 pt
\centerline { \title  Logarithmic Operators in Conformal Field Theory}
\baselineskip = 50 pt
\centerline {\subtitle V. Gurarie}
}
\vskip 20 pt
\centerline {\subtitle \sl Department of Physics}
\centerline {\subtitle \sl Princeton University}
\centerline {\subtitle \sl Princeton, NJ 08544}
{\baselineskip=100 pt
\centerline {\subtitle \bf Abstract }}
\vskip 20pt

{ Conformal field theories with correlation functions which have logarithmic
singularities are considered. It is shown that those singularities
imply the existence of additional operators in the theory which
together with ordinary primary operators form the basis of the Jordan
cell for
the operator $L_{0}$. An example of the field theory possessing such
correlation functions is given.
}

\vfill
\eject

\centerline {\subtitle 1. Introduction}

\vskip 15 pt

One of the basic assumptions of Conformal Field Theory is the
existence of the space of fields with the operator product
expansion. The coefficients of this expansion are supposed to be
powers of the distance and it can be used to construct all the
correlation functions.
For example,
the expansion of the four point correlation function of some
operators
in powers of the anharmonic ratio exactly corresponds to the
operator product expansion of those operators.
In many interesting cases
the Laurent series expansion exists. But it turns out there
are some cases when this expansion does not exist, which means the
operator product expansion as it is understood in Conformal Field
Theory must be modified.

 Let us
review some of the results obtained in [1] which we will need. One of them
was the existence of the so called primary operators
$ A_{k} ( z )$.
These operators transform under conformal transformations in a certain way,
namely when $ z \rightarrow f ( z )$,
$$ A_{k} ( z ) \rightarrow \left( { df \over dz
} \right)^{h_{k}} A_{k} \left( f \right) \eqno (1)$$
where $h_{k}$ is called the (right) dimension of the operator $A_{k}$.

The requirement of the correlation functions of the theory to be
conformaly invariant provides certain restrictions on the correlation
functions. In particular, the two point correlation function of two primary
operators must behave as a power of distance,
$$ < A_{k} (z) A_{k} (w) > = {const \over (z-w)^{2h_{k}}} \eqno (2) $$
The three point correlation functions are also fixed up to a constant.
As for the four point correlation functions of the primary operators,
they are fixed up to an arbitrary function of the so called anharmonic
ratio,
$$ < A_{k} (z_{1}) A_{k} (z_{2}) A_{k} (z_{3}) A_{k} (z_{4}) > =
{ 1 \over (z_{1}-z_{3})^{2h_{k}} (z_{2}-z_{4})^{2h_{k}} } F ( x )
\eqno (3) $$
where $F ( x ) $ is some function and $x$ is the anharmonic ratio $$
x={(z_{1}-z_{2})(z_{3}-z_{4}) \over (z_{1}-z_{3})(z_{2}-z_{4})} \eqno (4) $$

It is well-known that the expansion of $F ( x)$ in powers of $x$ corresponds
to the operator product expansion of Conformal Field Theory.
Conformal invariance requires the operator product expansion to have
the form
$$ B_{k}(z) B_{j}(w) = \sum_{n} {C_{n} \over (z-w)^{h_{k}+h_{j}-
h_{n}}} B_{n}(w) \eqno (5) $$
where $B$ are some (not only primary)  operators. On the other hand
if we substitute this expansion to (3), we will get exactly the
Laurent series expansion of the function $F(x)$.

The function $F(x)$ is not easy to determine. In most cases, only
its Laurent series expansion coefficients
can be calculated one after another, as was
shown in [1]. But in the cases of the so called degenerate operators
correlation functions are known to satisfy certain differential equations.
We can determine $F(x)$ by solving these equations. It turns out these
equations are the differential equations for the generalized hypergeometric
functions [2]. So, for the degenerate operator four point correlation
functions $F(x)$ are the generalized hypergeometric functions. In most cases,
these functions can be expanded in Laurent series around $x=0$. But sometimes
for the special values of the operator dimensions the functions $F(x)$ can
have logarithmic divergencies near $x=0$. In fact, the most general behavior
for $F(x)$ near $x=0$ is given by
$$F(x)=\sum_{n=0}^{\infty} a_{n} x^{n} + \log (x) \sum_{n=0}^{\infty}
b_{n} x^{n} + \log^{2} (x) \sum_{n=0}^{\infty} c_{n} x^{n} + \dots
\eqno (6)$$
Such a behavior is in explicit contradiction with the operator product
expansion (5).

We will have to introduce new operators in the operator product expansion,
operators which will have ``logarithmic'' behavior.
Ordinary primary operators are known to be the eigen vectors of the
$L_{0}$ operators,  and their eigen values are the
dimensions of these operators. It will be shown that those ``new''
operators, which I will call pseudo-operators, are the basis of the
Jordan cell for $L_{0}$.

As we will see, the logarithm-like singularities in the
correlation functions are quite a rare occasion in conformal
field theories. That's probably why they have not been studied before
even though the logarithms in the correlation functions of WZW models
have already appeared in [3].
They do not appear in unitary minimal models, and only nonminimal or
possibly nonunitary minimal models have a chance of possessing them.
However, such nonminimal models have recently emerged in the
studies of percolations [4] and polymers [5]. Also, the recent work on
conformal turbulence [6] has revived the interest in nonunitary
minimal models and provided us with new motivations to understand them
better.
We hope this paper contributes to that understanding.

\vskip 15pt

\centerline {\subtitle 2. Logarithmic pseudo operators}

\vskip 15pt

To begin with, we consider the case of the $c=-2$ model. It provides us
with the example of
the most simple correlation function with logarithmic behavior.
Really, let us take the operator $(1,2)$ with dimension
$-{1 \over 8}$. We will call it $\mu$ from now on. If we
try to compute the four point correlation function of the operator $\mu$
$$ < \mu (z_{1}) \mu (z_{2}) \mu (z_{3}) \mu (z_{4}) > =
(z_{1}-z_{3})^{1 \over 4} (z_{2}-z_{4})^{1 \over 4}
[x(1-x)]^{1 \over 4} F(x) \eqno (7) $$
we discover that the function $F(x)$ satisfies the following
differential equation as a consequence of the degeneracy of $\mu$
on the second level.
$$ x(1-x){d^{2}F(x) \over dx^{2}} +(1-2x) {dF(x) \over dx} - {1
\over 4} F(x) =
0 \eqno (8) $$
We want to find the asymptotic behavior of $F(x)$ as $x \rightarrow 0$.
Substituting $F(x) \sim x^{\alpha}$ and taking the leading term we get
$ \alpha^{2} =0$. Both two solutions of this equation are
$\alpha =0$. The standard interpretation of this fact by [1]
would be that both operators arising in the Operator Product Expansion
of $\mu(z) \mu(w)$ have dimension 0, so it seems they are the same unit
operators.
But this in fact is incorrect.
If these two $\alpha$ were different, we would really obtain the
asymptotic
behavior of two linear independent solution of (8) corresponding to the
contributions of two different primary operators and their descendants.
But since they are the same, simple analysis of (8) shows that
while
the first solution of (8) really behaves as $x^{0}$ the other solution
behaves as $\log (x)$.

One might think that one can simply discard the second solution of (8)
and do not worry about logarithms. But this is not true.  It is not
difficult to check that the first solution of (8) which is regular
at $x=0$ will have a logarithmic singularity at $x=1$. Therefore, it
is impossible
to get rid of the logarithms in
the OPE of the operators $\mu$.

Let us see how this works.
We look for the first solution of (8) in terms of the series
$$ \sum_{n=0}^{\infty} a_{n} x^{n} $$
After substituting it to (8) one can find the coefficients $a_{n}$ one after
another recursively. It gives us the first solution of (8) regular
at $x=0$. In fact, it is going to be the hypergeometric function
$F({1 \over 2}, {1 \over 2};1;x)$, but this is not important for us at
the moment. Now to look for the second solution of (8) we consider
the following candidate
$$ \sum_{n=0}^{\infty} b_{n} x^{n} \log (x) + c_{n} x^{n} $$
The first thing we see after substituting it to (8)
is $b_{n}=a_{n}$. Then $c_{n}$ can also be obtained one
after another.

Fortunately, the equation (8) is simple enough for its exact
solution to exist in a closed form.
It can be expressed in terms of the
elliptic integral
$$ G(x) \equiv \int_{\varphi=0}^{\pi \over 2} {d \varphi \over
\sqrt {1-x \sin ^2 (\varphi)}} \eqno (9) $$
and is given by
$$ F(x)=C_{1} G(x) + C_{2} G(1-x) $$
$C_{1}$ and $C_{2}$ are the arbitrary constants. $G(x)$ is regular at
$x=0$. The behavior of $G(1-x)$
can be understood if we use the following formula
$$ G(1-x)=G(x) \log (x) + H(x) \eqno (10) $$
where $H(x)$ is some regular at $x=0$ function. (10) of course follows
immediately from the above discussion of the solution properties.
A very important point which must be emphasized now is that the logarithmic
solution of (8) has the regular solution of (8) multiplied by the logarithm
plus some other regular function.

We are going to construct the operator product expansion of $\mu(z)
\mu(0) $ compatible with the solutions of (8). The first regular
solution of (8) implies the ordinary OPE
$$ \mu(z) \mu(0)|> = z^{ 1 \over 4} |I, z> \eqno (11) $$
where $|I, z>$ is the contribution of the identity operator and its
descendants (we will work with the states instead of the operators for a
moment; $|>$ is the vacuum). The second solution of (8) corresponds to a new
OPE
$$ \mu(z) \mu(0)|> = z^{1 \over 4}  (\log (z) |I , z> + |I_{1},z>)
\eqno (12) $$ where $|I_{1}, z> $ denotes the contribution of the
new pseudo-operator with unusual properties.

The dilatation transformation performed on (12) hints we must have
$$ L_{0} I_{1} = I $$
This can be done more rigorously by the method developed in [1].
Namely let us expand $ |I,z>$ and $|I_{1}, z >$ in series
$$ |I, z> = \sum_{n=0}^{\infty} z^{n} |I, n> $$
and
$$ |I_{1}, z> = \sum_{n=0}^{\infty} z^{n} |I_{1}, n> $$
substitute these series to (12) and apply $L_{k}$ to both sides of the
equality obtained. Then the standard procedure usually used to
determine coefficients in the conformal block and including the
commuting of $L_{k}$ with $\mu(z)$ and using
$$ [L_{k}, A(z)] = \left( z^{k+1} { \partial \over \partial z}+ (k+1) h_{A}
z^k \right)  A(z) $$
for any primary operator $A(z)$
will give
$$ \matrix { L_{0} |I,n> =n |I,n> \cr L_{0} |I_{1},n> = |I,n> + n|I_{1},n>}
 \eqno (13)
$$
$$ \matrix {L_{k}|I, n+k> = (n+(k-1)h_{\mu}) |I, n> \cr
   L_{k}|I_{1}, n+k> = |I, n> + (n+(k-1)h_{\mu}) |I_{1},n>}
   \eqno (14) $$
$ h_{\mu}= - {1 \over 8}$ is the dimension of $\mu$, $k >0$ in (14).

These equations accumulate all the properties of the operator $I_{1}$.
First we see that
$$ L_{0} |I_{1}, 0>=|I,0> \eqno(15) $$
as we expected. So, the operators $I_{1}$ and $I$ are the basis of the
Jordan cell for $L_{0}$. Secondly, we can explicitly write
$$ |I_{1},1> = \beta_{1} L_{-1} |I_{1}, 0>,
    $$
$$ |I_{1},2> = \beta_{2} L_{-2} (a|I,0> + |I_{1}, 0 > ) +
               \beta_{11} (L_{-1})^{2} ( b|I,0>+|I_{1},0>) $$
and so on, then the equations (14) help us to determine all $\beta$'s,
$a$, $b$,
etc.\footnote{${}^{\dagger}$} {
 We must include descendants of $|I,0>$ into
$|I_{1},k>$ for consistency. Let us note that it doesn't contradict (13).
We do not include $L_{-1} |I,0>$ to $|I_{1}, 1>$ as $L_{-1} |I,0> =0$.}

It is also useful to see how conformal bootstrap can be solved in this case.
Following [2] we must construct a single-valued function of $x$ and
$\bar x$ out of the solutions of (8) which will give us  the ``full''
correlation function, with its holomorphic and antiholomorphic
dependence. Using (10) we can check that the answer is
$$ (z_{1}-z_{3})^{1 \over 4} (z_{2}-z_{4})^{1 \over 4} [x(1-x)]^{1 \over
4}
\overline {(z_{1}-z_{3})^{1 \over 4} (z_{2}-z_{4})^{1 \over 4}
[x(1-x)]^{1 \over 4}}
[G(x)G(1-\bar x)+ G(\bar x) G(1-x)] \eqno (16) $$
This answer is different from the ordinary case of ``normal'' operators,
the usual answer would be $\sum_{k} G_{k}(z) G_{k}(\bar z)$ while here
we have something like $G_{1}(z)G_{2}(\bar z) + G_{2}(z)G_{1}(\bar z)$,
$G_{k}$ are the linear independent solutions for the correlation function
differential equation. We interpret this fact by noting that the
``full'' operators in the theory must have a normal holomorphic and
a pseudo antiholomorphic dimension or vice versa.

Now we can generalize the above scheme. Consider the model of the
conformal field theory where there are two operators OPE of which, found
according to the rules of [1], contains at least two operators with
the same dimensions. As we know now, that means these rules are no
longer valid and in fact these two operators are the basis of the Jordan
cell for $L_{0}$. So, let
$$ A(z) B(0) = z^{h_{C}-h_{A}-h_{B}} \{
C_{1} + \dots + \log(z) (C + \dots) \} \eqno (17) $$
After repeating the same arguments we arrive at the following
formulae

$$ \matrix
   { L_{k}|C, n+k> = (h_{C}+n-h_{B}+kh_{A})|C,n> \cr
     L_{k}|C_{1}, n+k> = |C,n> +(h_{C}+n-h_{B}+kh_{A})|C_{1},n> }
\eqno (18) $$
$$ \matrix {
     L_{0}|C, n> = (h_{C}+n) |C,n> \cr
     L_{0}|C_{1}, n> = |C,n> + (h_{C}+n)|C_{1},n>
} \eqno (19) $$
which are, not surprisingly,
the straightforward generalizations of (13) and (14).
Using (19) and (19) we can express $|C,n>$ and $|C_{1},n>$ in terms
of the descendants of $C$ and $C_{1}$ the same way as we have done
already for the case
 of $I$ and $I_{1}$. The answer for the first level is
$$\matrix {
|C,1>={ h_{C}+h_{A}-h_{B} \over 2 h_{C} } L_{-1} |C,0> \cr
|C_{1},1>={ h_{C}+h_{A}-h_{B} \over 2 h_{C} } L_{-1} |C_{1},0> +
   { h_{B}-h_{A} \over 2 h_{C}^{2} } L_{-1} |C, 0>
} \eqno (20) $$

It may be interesting to try to find correlation function containing the
operator $C_{1}$. We can proceed here in many different ways. For
example, we can take the correlation function
$$ < A(z_{1}) B(z_{2}) A(z_{3}) B(z_{4}) > $$
and
expand the product $A(z_{3}) B(z_{4})$ according to (17). On the
other hand, we know this function is going to behave like
$$ {1 \over (z_{1}-z_{3})^{h_{A}}} {1 \over (z_{2}-z_{4})^{h_{B}}}
{1 \over [x (1-x)]^{h_{A}+h_{B}-h_{C}}} [ \log(x) (a_{0}+a_{1}x+\dots )  +
b_{0}+b_{1}x+\dots ] $$
So, if we take this expression, substitute the value for $x$ in terms of
$z_{1}$, $z_{2}$, $\dots$,  and extract from it a coefficient multiplying
 $(z_{3}-z_{4})^{h_{C}-h_{A}-h_{B}}$, we will get exactly

$$ <A(z_{1}) B(z_{2}) C_{1} (z_{3}) > = <A(z_{1}) B(z_{2}) C(z_{3})>
\left[ \log { z_{1}-z_{2} \over (z_{1}-z_{3})(z_{2}-z_{3}) }
+ \lambda \right] \eqno (21) $$
This function turns out to be invariant with respect
to the translation, dilatations,
rotations and special conformal transformations.\footnote{${}^{\dagger}$}
{ When checking this statement, one has to keep in mind that the
transformation law for $C_{1}$ is given by $\delta C_{1}(z) =
\epsilon'(z)[h_{C} C_{1}(z) + C(z)] +\epsilon(z) C_{1}'(z) $ }
The appearance of the arbitrary parameter $\lambda$ is due to the fact
that we can always replace $C_{1} \rightarrow C_{1}+\lambda C$.

We can continue the expansion of (21) to get two point correlation functions.
which turn out to be

$$ \matrix {
<C_{1}(z) C_{1}(w)> =  - {2 \over (z -w)^{2h_{C}}} [ \log (z-w)
+ \lambda' ] \cr
<C(z) C_{1}(w)> = {1 \over (z-w)^{2h_{C}}}
}
\eqno (22) $$

It is not possible to determine $\lambda$'s without knowing the
structure constants of the operator algebra.

Another way to find the correlation functions is to study the operator
 product expansion of $C(z) C_{1}(w)$, $A(z) C_{1}(w)$ etc. by the
method used to derive (13) and (14). Let us again consider for simplicity
the case of $c=-2$. We want, for example, to find the operator
product expansion of $\mu(z) I_{1}(w)$. To do that, first we denote
$$\mu(z)I(0)|>=
|\mu,z>_{1} \eqno(23)$$ The next step will be to assume that
$$\mu(z) I_{1}(0) |>=
\log(z)|\mu,z>_{2}+|\mu,z>_{3} \eqno(24)$$ We denote here the blocks by
$|\mu,z>_{1}$, $|\mu,z>_{2}$, and $|\mu,z>_{3}$ to emphasize that the
coefficients of the expansions in terms of the descendants of $\mu$ in
these blocks may be different. And
the last step is to apply $L_{k}, k\ge 0$ to both sides of this equation.
Taking into account the unusual property of $I_{1}$, namely that
$L_{0}I_{1}=I$, we get
$$ \matrix{ \left( z { \partial  \over \partial z } + h_{\mu} \right) \left(
\log(z)|\mu,z>_{2}+|\mu,z>_{3} \right)
+ |\mu,z>_{1} = \cr
(\log(z) L_{0} |\mu,z>_{2} + L_{0}|\mu,z>_{3}) } \eqno (25) $$
$$ \matrix{ \left( z^{k+1}{ \partial  \over \partial z }
+ (k+1) h_{\mu} z^{k}\right) \left(
\log(z)|\mu,z>_{2}+|\mu,z>_{3} \right)
 = \cr
\log(z) L_{k} |\mu,z>_{2} + L_{k}|\mu,z>_{3} } \eqno(26) $$
$k>0$

It is clear from (25) and (26) is $|\mu,z>_{2}= -|\mu,z>_{1}$.
Then the structure of $|\mu,z>_{3}$
can also be determined.
So, we proved that $$\mu(z)I_{1}(0)|>=-\log(z)\mu(z)I(0)|>+|\mu,z>_{3}
\eqno(27) $$
$|\mu,z>_{3}$ is a conformal block different from the one arising in
the expansion of $\mu(z) I(0)|>$.

Similar technique can be applied to find any OPE in the theory with
pseudo-operators. These OPE will be compatible with the correlation
functions (21) and (22).

Our construction can be naturally generalized to the case of three or
more operators forming the basis of the Jordan cell for $L_{0}$.
Then we have  $m$ operators $C_{n}$ such that
$$ \matrix {L_{0}C_{n}=h_{C}C_{n}+C_{n-1}, n>0 \cr
           L_{0}C_{0}=h_{C}C_{0} } \eqno (28) $$
and they enter the OPE in the following form

$$ \matrix {A(z)B(0)=z^{h_{C}-h_{B}-h_{A}} \{C_{0}+\dots\} \cr
            A(z)B(0)=z^{h_{C}-h_{B}-h_{A}} \log(z)\{C_{0}+\dots\} +
 z^{h_{C}-h_{B}-h_{A}} \{C_{1}+\dots\} \cr
\hbox to 3 in {\dotfill } \cr
A(z)B(0)= z^{h_{C}-h_{B}-h_{A}}
\sum_{n} \log^{n}(x)\{C_{m-n}+\dots\}} \eqno(29) $$

{}From the above discussion it is clear we must include logarithmic
operators in the theory if it possesses at least two operators the
product of which when expanded according to the fusion rules of [1]
contains the contribution of at least two operators with the same
dimension. It can be proved such a case never occurs in the minimal
models. But it always occurs in any non-minimal model. So, non-minimal
models are in fact richer than it was thought they were.

\vskip 15pt

\centerline {\subtitle 3. Other representations of  the Virasoro algebra.}

\vskip 15pt

There is another case when logarithmic correlation functions occur in
the conformal field theory. Namely, it is the case when there are
operators
in the theory the OPE of which contains operators dimensions of which
differ by the integer number. Really, if this is the case, then the
equation for the function $F(x)$  defined analogously to $F(x)$ in
(7) possesses two
asymptotic solutions, $x^{0}$ and $x^{n}$, $n$ is some positive integer
number.  Then there are two possibilities. Either the function
$$ F(x)=\sum_{m=0}^{n-1} a_{m} x^{m} \eqno (30)$$
with the appropriate choice of $a_{m}$
is an exact solution of the corresponding differential equation or
the two linear independent solutions of this equation are
$$ F(x)=\sum_{m=n}^{\infty}b_{m-n}x^{m}\eqno (31)   $$
and
$$ F(x)=\sum_{m=0}^{\infty}c_{m}x^{m}+\log(x)\sum_{m=n}^{\infty}b_{m-n}x^{m}
\eqno(32)$$

Here we again have to introduce the logarithmic operators. But their
behavior will be more complicated. The solution (31) corresponds
to the contribution of the ordinary primary operator while
the solution (32) corresponds to the contribution of the same operator
multiplied by $\log(x)$ plus the contribution of  another pseudooperator
with even more unusual properties than the ones considered in section
2.

Really, this operator must have dimension smaller than the dimension
of its ordinary partner, but still the operator $L_{0}$ must again
be nondiagonal in the same fashion as it was before.
If we suppose that simply $L_{0}|C_{1},m>=|C,m-n>+(h_{C_{1}}+m)|C_{1},m>$ if
$m\ge n$ and $L_{0}|C_{1},m>=(h_{C_{1}}+m)|C_{1},m>$ otherwise, we
immediately obtain a contradiction. By applying $L_{-1}$ to
$$ L_{0}|C_{1},n-1>=(n-1+h_{C_{1}})L_{-1}|C_{1},n-1>$$
we get
$$ L_{0}L_{-1}|C_{1},n-1>=(n+h_{C})L_{-1}|C_{1},n-1> $$
which is difficult to satisfy in view of
$$ L_{0}|C_{1},n>=|C,0>+(h_{C_{1}}+n)|C_{1},n>$$.

We have  to use a more complicated construction. Let us make the
operator $C_{1}$ have the dimension equal to that of $C$ with the same
Jordan-like philosophy, $L_{0}|C_{1},m>=|C,n>+(m+h_{C})|C,m>$ but also
let us make the operator $C_{1}$ to be ``even less'' primary, that is
$L_{n}|C_{1}>=|C^{1}>$ where $|C^{1}>$ is a primary operator of the
dimension $h_{C}-n$. Then the OPE corresponding to (31) and (32)
are
$$ A(z)B(0)= z^{h_{C}-h_{A}-h_{B}}(C(0)+\dots ) \eqno (33) $$
$$ A(z)B(0)= z^{h_{C}-h_{A}-h_{B}}\{\log(z) (C(0)+\dots)+
(C_{1}+\dots)+z^{-n}(C^{1}+\dots)\} \eqno (34) $$
Then we can work out in the same way all the coefficients in the
operator product expansion of different operators.

The simplest example (and unfortunately the only one
 known to me) when all discussed above is the case can be
found in the same $c=-2$ theory. If we compute the correlation function
of the operators (1,2) and (2,2) of this theory we will obtain
$$<A_{(1,2)}(z_{1})A_{(2,2)}(z_{2})A_{(1,2)}(z_{3})A_{(2,2)}(z_{4})> =
(z_{1}-z_{3})^{1 \over 4} (z_{2}-z_{4})^{-{3 \over 4}}
[x(1-x)]^{3 \over 4} F(x) \eqno (35) $$
where the function $F(x)$ turns out to be the first derivative of
(9). So, its expansion around the logarithmic point is given by the
derivative of (10),
$$ F(x) = G'(x) \log(x)+{G(x) \over x} + H'(x) $$
which is of the same form as (32). We see that the construction
of OPE of $A_{(1,2)}(z)A_{(2,2)}(w)$ includes one normal operator
$W$ with the dimension $1$, one ``pseudo'' operator $W_{1}$ with the
dimension 1 and the property $L_{0}W_{1}=W+W_{1}$ and one primary
operator $W^{1}$ with the dimension 0 and the property
$L_{-1}W_{1}=W^{1}$. The behavior of $W^{1}$ will almost always coincide
with that of the unit operator, it has the same dimension after all and
all $L_{k} W^{1}=0$ for $k>0$.

We can proceed to find the OPE of these operators.
It will be the same procedure as the one described in section 2.
For example, the expansion of $A_{(1,2)}(z) W_{1}(0)$ will be given by

$$ A_{(1,2)}(z) W_{1}(0) |> = - z^{-{ 5 \over 4}} (\log(z) |A_{(2,2)},z>_{1} +
|A_{(2,2)},z>_{2}) + z^{-1} |A_{(1,2)},z> $$
where the first two terms on the right hand side
are constructed analogously to (27) and the third term satisfies the
equations

$$ \matrix  {
z^{-1} L_{1} |A_{(1,2)},z> = \left( z^{2} { \partial \over \partial z}+ 2 h_{A}
z \right) z^{-1}|A_{(1,2)},z> +  A_{(1,2)}(z) W^{1}(0) |>
\cr
z^{-1} L_{2} |A_{(1,2)},z> = \left( z^{3} { \partial \over \partial z}+
3 h_{A} z^{2} \right) z^{-1} |A_{(1,2)},z> } $$

The interesting question is whether the case of such logarithmic
operators occurs in the minimal models. I do not know yet the answer
to this question. All minimal models I studied explicitly so far
have   been the examples of (30). Nevertheless, the possibility of
nonunitary minimal model with correlation functions (31) and (32) cannot
 be ruled out
at this point.

\vskip 15 pt

\centerline {\subtitle 4. Conclusions}

\vskip 15 pt

A physical example
of the theory with logarithmic operators can be provided by the
free ghost model with the action
$$ S \sim \int \partial \theta \bar \partial \bar \theta \  d^{2} z \eqno (36)
$$
 where $\theta$ and $\bar \theta $ are the anticommuting variables.
This theory is analogous to the $\eta$, $\xi$ system studied in [5]
in connection with polymers.

It is shown in the appendix that
the central charge of this theory is equal to
$c=-2$ and its physics can be described by the
non-minimal model extensively discussed in this paper.
In particular, all the operators $(n,1)$ with the dimensions
${n^{2}-n \over 2}$ correspond to $$\partial \theta \partial^2 \theta
\dots \partial^{(n-1)} \theta \eqno(37a) $$
or
$$\partial \bar \theta \partial^2 \bar \theta
\dots \partial^{(n-1)} \bar \theta \eqno (37b) $$
 Then the
primary operators $(1,2)$ and $(2,2)$
will be twist operators of this theory and their correlation functions
will have logarithmic singularities. This implies there are also
logarithmic pseudooperators in this theory as was shown in the numerous
examples in this paper.

It was already mentioned in the Introduction that according to Cardy [4]
the correlation function desribing percolations should be taken from
the nonminimal
$c=0$ theory. Even though this function by itself does not possess
logarithmic singularities, other functions of this theory do. The OPE
of the operators underlying this theory will also include logarithmic
pseudooperators.

So, we have shown  some of conformal field theories necessarily possess
some other operators in addition to just primary. Unfortunately, the
task of classifying all such theories has
not been accomplished here. The main difficulty is to find a way to
distinguish differential equation with a polynomial solution (30)
from that with a logarithmic solution (32). One of the ways to overcome
that difficulty is to use explicit formulas for the structure constants of
conformal field theory given in [2]. It turns out those formulas give
infinity in the case of logarithmic conformal field theories,
technically due to
gamma functions of negative integers. It would be interesting to
carry out such an analysis of those structure constants. It will not
be an easy task either, since the structure constants are given by
long and ugly expressions.

\vskip 20 pt

\centerline { \subtitle Appendix}

\vskip 15pt

 We are going to study the free ghost model described by the action
(36) to show that it is described by the $c=-2$ minimal model.

With the appropriate normalization of the action, the correlation
function is given by
$$ < \partial \theta (z) \partial \bar \theta  (w) > = - { 1
\over ( z - w)^2 } \eqno (a.1) $$

Then the stress-energy tensor is given by
$$ T(z) = :\partial \theta (z) \partial \bar \theta ( z ) : \eqno (a.2)
$$
and the central charge is equal to $c=-2$.

Let us consider the twist fields of this theory. Let
$$ \partial \theta (z) = \sum_{n} \theta_{n} z^{-n-1},
\partial \bar \theta ( z)
= \sum_{n} \bar \theta_{n} z^{-n-1}$$
Then
$$ \{ \theta_{n} \bar \theta_{m} \} = - n \delta_{n+m,0} \eqno (a.3)$$
$\{ \}$ means anticommutator.
Then we assume $n, m \in {\bf Z} + {1 \over 2}$. To study the properties
of the twist fields obtained we have to impose the OPE of $\partial
\theta$ with those fields. If we assume $\partial \theta (z) | \sigma>
\sim \sqrt {z}$, $|\sigma>$ is the state assosiated with
the corresponding
twist field, then
$$ \theta_{n} |\sigma> =0, n \geq  - {1 \over 2} $$

It gives rise to

$$ <\sigma | \partial \theta (z) \partial \bar \theta (w) | \sigma > =
- \sum_{n={3 \over 2}, {5 \over 2}, \dots} n w^{n-1} z^{-n-1} =
 - \left( {1 \over (z-w)^2} + {1 \over 2 z (z-w) } \right) \sqrt {w \over
z} \eqno (a.4) $$

Now if $w = z+ \epsilon$ then
$$ <\sigma | \partial \theta (z) \partial \bar \theta (z + \epsilon
) | \sigma > \approx
-{1 \over \epsilon^2} + {3 \over 8 z^2} $$
which means $$L_{0}|\sigma> = {3 \over 8 } | \sigma>$$

Quite analogously,  $\partial \theta(z) | \mu> \sim \sqrt {1 \over
z}$ gives
$$ <\mu| \partial \theta (z) \partial \bar \theta (w) |\mu> = -
{ z + w \over 2 (z-w)^2 \sqrt{ z w} } \eqno (a.5) $$ which gives
$$ L_{0}|\mu>= - {1 \over 8 } |\mu>$$

The dimensions of the fields $\mu$, $\sigma$ and $\partial \theta$
 coincide with the
dimensions of the fields $A_{(1,2)}$, $A_{(2,2)}$ and $A_{(2,1)}$
of the conformal model
$c=-2$. It is alluring to conclude they are the same fields. To prove
it,
all we have to do is to compare the correlation functions of those
fields.

It is straightforward to check that the correlation functions of the
operators $A_{(2,1)}$ are the same as those of $\partial \theta$ or $
\partial \bar \theta$. Actually, to make the correspondence even
more clear, the following construction is useful. We can identify
$$ A_{(2,1)} \sim \sigma_{x} \partial \theta + \sigma_{y} \partial \bar
\theta \eqno (a.6)$$
where $\sigma_{x}$ and $\sigma_{y}$ are Pauli matrices.
The OPE of those operators will be given by
$$ ( \sigma_{x} \partial \theta(z) + \sigma_{y} \partial \bar
\theta(z) )  ( \sigma_{x} \partial \theta(0) + \sigma_{y} \partial \bar
\theta(0) ) = \sigma_{z} ( I + \dots) + ( \partial^2 \theta(0) \partial
\theta(0)+ \partial^2 \bar \theta(0) \partial \bar \theta (0) + \dots
) \eqno (a.7) $$

The first brackets symbolize the contribution of the unit operator,
while the second ones show the contribution of the operator $A_{(3,1)}$
with the dimension $3$. This example shows how the series of the
operators (37) discussed in the Conclusions section has appeared.

It is no more difficult to calculate the correlation functions with
$A_{(1,2)}$ and $A_{(2,2)}$. We get
$$ \lim_{z_{4} \rightarrow 0 \atop z_{2} \rightarrow \infty
} z_{2}^{- {1 \over 4}} <A_{(2,1)} (z_{1}) A_{(1,2)} (z_{2}) A_{(2,1)}
 (z_{3}) A_{(1,2)} (z_{4})> = { 1 \over (z_{1} - z_{3})^2} {
z_{1}+z_{3} \over \sqrt {z_{1} z_{3}} } \eqno (a.8)$$
which coincides with $(a.5)$ up to a constant.

 The analogous calculations
prove the same for the fields $A_{(2,2)}$ and $\sigma$.

Now we are allowed to calculate the correlation functions $<\mu \mu \mu
\mu>$ using the rules of conformal field theory. As we already know those
functions will have logarithmic singularities which means there
are logarithmic operators in the OPE of $\mu$'s and $\sigma$'s. Let us
again stress that even though we do not know how to construct them
explicitly we can always draw the following logical line: free ghosts
$\rightarrow$ twisted correlation functions $\rightarrow$ twist fields
$\rightarrow$ their correlation functions $\rightarrow$ logarithmic
pseudooperators.

\vskip 15 pt

\centerline {\subtitle Acknowledgements.}

\vskip 15 pt

The author is grateful to A.M.Polyakov for encouragement during
this
work and to A. Matytsin and D. Polyakov for stimulating discussions.

\vskip 30pt

\noindent
 {\bf References}

\noindent
[1] A. Belavin, A. Polyakov and A. Zamolodchikov, {\sl Nucl. Phys.
\bf
B241}, 33 (1984).

\noindent
[2] Vl.S. Dotsenko and V.A. Fateev  {\sl Nucl. Phys. \bf B240}, 312
(1984).

\noindent
[3] L. Rozansky, H. Saleur {\sl Nucl. Phys. \bf B376}, 461
(1992)

\noindent
[4] J. Cardy, Preprint UCSBTH-91-56   (1991)

\noindent
[5]  H. Saleur, Preprint YCTP-P38-91  (1991)

\noindent
[6]  A.Polyakov, Preprint PUPT-1369 (1992)

\bye